\documentclass[journal]{IEEEtran}

\usepackage{enumitem}

\usepackage{cite}

\usepackage{amsmath}
\usepackage{algorithm}
\usepackage{algorithmic}
\usepackage{amssymb}
\usepackage{array}
\usepackage{setspace}
\usepackage{blkarray}
\usepackage{environ}
\usepackage{multirow}
\usepackage{url}
\usepackage{etoolbox}
\usepackage{float}
\usepackage{nomencl}
\usepackage{marvosym}
\usepackage{soul}

\usepackage{epsfig}
\usepackage{bm}
\usepackage{latexsym}
\usepackage{romannum}
\usepackage{amsfonts}
\usepackage[pdftex]{hyperref}
\graphicspath{{./pic/}} \DeclareGraphicsExtensions{.pdf,.jpg,.png}
\hypersetup{colorlinks=true, allcolors=black, linkcolor=black, citecolor=black }
\usepackage{mathtools}

\usepackage{mdwlist}
\usepackage{graphicx}

\usepackage{footnote}
\makesavenoteenv{tabular}
\usepackage{tabulary}
\usepackage[table]{xcolor}
\newcolumntype{P}[1]{>{\centering\arraybackslash}p{#1}}
\newcolumntype{M}[1]{>{\centering\arraybackslash}m{#1}}
\usepackage[caption=false, font=footnotesize]{subfig}

\begin{document}
%
\title{Optimal Placement of PV Smart Inverters with Volt-VAr Control in Electric Distribution Systems }

\author{Mengxi Chen,~\IEEEmembership{Student Member,~IEEE,}
        Shanshan Ma,~\IEEEmembership{Member,~IEEE,}
        Zahra Soltani,~\IEEEmembership{Student Member,~IEEE,}\\
        Raja Ayyanar,~\IEEEmembership{Senior Member,~IEEE,}
Vijay Vittal,~\IEEEmembership{Life Fellow,~IEEE,}
	Mojdeh Khorsand,~\IEEEmembership{Member,~IEEE}
\thanks{The work is funded by the Department of Energy (DOE) Advanced Research Projects Agency-Energy (ARPA-E) under OPEN 2018 program.}
\thanks{M. Chen, S. Ma, Z. Soltani,R. Ayyanar, V. Vittal,  and M. Khorsand are with the School of Electrical, Computer, and Energy Engineering, Arizona State University, Tempe, AZ 85281, USA (Email: Mengxi.Chen@asu.edu, Shanshan.Ma@asu.edu, rayyanar@asu.edu, Vijay.Vittal@asu.edu, zsoltani@asu.edu; mabdikho@asu.edu).}
}

\maketitle

\begin{abstract}
The high R/X ratio of typical distribution systems makes the system voltage vulnerable to the active power injection from distributed energy resources (DERs). Moreover, the intermittent and uncertain nature of the DER generation brings new challenges to the voltage control. This paper proposes a two-stage stochastic optimization strategy to optimally place the PV smart inverters with Volt-VAr capability for distribution systems with high photovoltaic (PV) penetration to mitigate voltage violation issues. The proposed optimization strategy enables a planning-stage guide for upgrading the existing PV inverters to smart inverters with Volt-VAr capability while considering the operation-stage characteristics of the Volt-VAr control. One advantage of this planning strategy is that it utilizes the local control capability of the smart inverter that requires no communication, thus avoiding issues related to communication delays and failures. Another advantage is that the Volt-VAr control characteristic is internally integrated into the optimization model as a set of constraints, making placement decisions more accurate. The objective of the optimization is to minimize the upgrading cost and the number of the smart inverters required while maintaining the voltage profile within the acceptable range. Case studies on an actual 12.47kV, 9km long Arizona utility feeder have been conducted using OpenDSS to validate the effectiveness of the proposed placement strategy in both static and dynamic simulations.
\end{abstract}


\section*{Nomenclature} 
\small
\begin{flushleft}
\textit{Sets and Indices}
\end{flushleft}
\begin{description}[style=multiline,leftmargin=2cm]
\item[$\varphi$] Set of phase indices $p$
\item[$\varphi(\ell)$] Subset of phase at distribution line $\ell$	
\item[$\Omega_\mathcal{B}$] Set of bus indices $(i,p)$
\item[$\Omega_\mathcal{D}\!\subset\!\Omega_\mathcal{B}$] Subset of bus with load indices $(d,p)$
\item[$\Omega_\mathcal{L}$] Set of distribution lines indices $(\ell,p)$
\item[${\mathcal{L}_O}(i)$] Subset of distribution lines $(\ell,p)$ originating at bus $i$
\item[${\mathcal{L}_E}(i)$] Subset of distribution lines $(\ell,p)$ ending at bus $i$
\item[$\ell_o,\ell_e$] From-node and to-node indices of distribution line $\ell$
\item[$\Omega_\mathcal{PV}\!\subset\!\Omega_\mathcal{B}$] Subset of bus with PV indices $(g,p)$
\item[$\Omega_{\mathcal{PV}_{\!v\!v}}\!\!\subset\!\Omega_\mathcal{PV}$] Subset of bus with PV candidates for smart inverters indices $(g,p)$
\item[$\Omega_{T\!R}$] Set of transformers indices $(r,p)$
\item[$\Omega_{sub}$] Set of substation indice $(n,p)$
\item[$\mathcal{S}$] Set of scenarios indices $s$ 
\end{description}
\begin{flushleft}
	\textit{Parameters}
\end{flushleft}
\begin{description}[style=multiline,leftmargin=2.1cm]
\item[$R_{\ell,p,m}$] Self resistance ($p=m$) and mutual resistance of line $\ell$ between phases $p$ and $m$ ($p\neq m$) 
\item[$X_{\ell,p,m}$] Self reactance ($p=m$) and mutual reactance of line $\ell$ between phases $p$ and $m$ ($p\neq m$)
\item[$Z_{\ell,p,m}$] Self impedance ($p=m$) and mutual impedance of line $\ell$ between phases $p$ and $m$ ($p\neq m$)
\item[$y_{\ell,m,k}$] Susceptance of line $\ell$ between phases $m$ and $k$
\item[$\bar{P}^{pv,s}_{g,p}$] The maximum power point (MPP) of PV at bus $g$ phase $p$ at scenario $s$
\item[$\bar{P}^{pv,\max}_{g,p}$] The rating active power value of PV inverter at bus $g$ phase $p$
\item[$\bar{Q}^{pv,\max}_{g,p}$] The rating reactive power value of PV inverter at bus $g$ phase $p$
\item[$\bar{S}^{pv}_{g,p}$] The rating apparent power value of PV inverter at bus $g$ phase $p$
\item[$\bar{V}^1_g,\dots,\!\bar{V}^6_g$] Voltage magnitude breakpoints of the linear piece-wise Q-V curve
\item[$V^{\min}\!/V^{\max}$] The minimum/maximum voltage magnitude for the normal operation
\item[$\bar{V}^{r,s}_{n,p}/\bar{V}^{im,s}_{n,p}$] Real/imaginary part of voltage measurement at substation before enabling Volt-VAr control $n$ phase $p$ at scenario $s$
\item[$p_r(s)$] The scenario probability 
\vspace{3pt}
\item[$w_c, w_o^s, w_v^s$] The weight coefficients in the objective function 
\item[$P^{G,s}_{n,p}$] Injection from substation $n$ at phase $p$ at scenario $s$
\item[$P^{Tr,s}_{r,p}$] No-load loss of transformer $r$ at phase $p$ at scenario $s$ 
\item[$P^{D,s}_{d,p}$] Active power of load at bus $d$ phase $p$ at scenario $s$ 
\item[$Q^{D,s}_{d,p}$] Active power of load at bus $d$ phase $p$ at scenario $s$
\item[$M$] A large positive number 
\item[$P\!F_{\min}$] The minimum power factor of PV smart inverter
\end{description}
\begin{flushleft}
\textit{Variables}
\end{flushleft}
\begin{description}[style=multiline,leftmargin=2cm]
\item[$x^{pv}_{g,p}$]  Binary variable to decide whether to place an PV smart inverter or not at bus $g$ phase $p$
\item[$x^{{pv}_{on},s}_{g,p}$]  Binary variable to enable PV Volt-VAr control function at bus $g$ phase $p$ at scenario $s$
\item[$x^{{pv}_{o\!f\!f},s}_{g,p}$]  Binary variable to disable PV Volt-VAr control function at bus $g$ phase $p$ scenario $s$
\item[$P^{pv,s}_{g,p}$] Active power output of PV at bus $g$ phase $p$ at scenario $s$
\item[$Q^{pv,s}_{g,p}$] Reactive power output of PV at bus $g$ phase $p$ at scenario $s$
\item[$Q^{qv,s}_{g,p}$] Reactive power output of PV following Q-V curve at bus $g$ phase $p$ at scenario $s$
\item[$I^{r,s}_{\ell,p}$] Real part of current flow at line $\ell$ phase $p$ at scenario $s$
\item[$I^{im,s}_{\ell,p}$] Imaginary part of current flow at line $\ell$ phase $p$ at scenario $s$
\item[$I^{in,r,s}_{i,p}$] Real part of current injection at bus $i$ phase $p$ at scenario $s$
\item[$I^{in,im,s}_{i,p}$] Imaginary part of current injection at bus $i$ phase $p$ at scenario $s$
\item[$V^{r,s}_{i,p}$] Real part of voltage at bus $i$ phase $p$ at scenario $s$
\item[$V^{im,s}_{i,p}$] Imaginary part of voltage at bus $i$ phase $p$ at scenario $s$
\item[$V^{m,s}_{i,p}$] Voltage magnitude at bus $i$ phase $p$ at scenario $s$
\item[$\phi(x^{pv}\!,s)$] Operation cost of the second stage at scenario $s$ given $x^{pv}$
\end{description}
\normalsize
\section{Introduction}

\IEEEPARstart{T}{he} rapid growth of distributed energy resources (DERs), especially the solar photovoltaic (PV) generators in distribution systems, is transforming the systems from passive networks to active ones\cite{Bedawy2020}. The large R/X ratio of the active distribution feeders makes the voltages sensitive to the intermittent active power injection from the PVs, which may lead to unexpected voltage violations and voltage fluctuations \cite{Kyriaki2017}.

There are four types of voltage control strategies applied in active distribution systems: local control, distributed control, decentralized control, and centralized control \cite{Sun2019}. The local control is an autonomous control strategy that does not require any communication among different controllers \cite{Alyami2014, Baker2018}. The centralized control can provide more flexibility by allowing the optimal utilization of the control devices among the entire system. However, a robust communication system is required to provide the real-time measurements for the central controller \cite{Juamperez2014, Rogers2010}. Distributed control is a strategy that does not need a central controller; only the communication between the neighboring controllers is required \cite{5_3Robbins2013}. The decentralized control coordinates various control components to optimize the operation for a specific area, which means the system can be divided into different ``centralized'' zones \cite{Gao2018}. These different control strategies are all based on the system operation stage with possible use of optimization techniques to coordinate the operation of different existing devices. However, the existing control devices in some distribution systems may not be adequate for handling the voltage issues caused by the high penetration of the PVs. Even worse, the bidirectional power flow of the active distribution network can mislead traditional control devices such as the tap changers and cause unexpected regulating or protecting actions \cite{Bedawy2020}.

To overcome these voltage management complexities, it is necessary to adopt an effective control strategy in the operational stage and develop an effective planning strategy that takes into account the operation of active voltage control. Many approaches on distribution system planning associated with active voltage management have been investigated\cite{28Ye2019,39Qian2019,29Asano2020}.Those planning strategies mainly focused on increasing the penetration level of the DERs. However, the system voltage regulation flexibility is limited by the capability of conventional control devices and the communication system. To further mitigate voltage violation issues, the interconnection standard IEEE 1547-2003 has been thoroughly revised and published as 1547-2018 to allow smart inverter-based generators to participate in the distribution feeder voltage regulation by providing sufficient active and reactive power support \cite{IEEE15472018}. This amendment enables the DERs with smart inverters to control and optimize the local voltage by injecting or absorbing reactive power based on the local operation condition. For a distribution system with a high penetration level of residential PVs, selectively upgrading existing PV inverters to the smart inverters with local voltage management capability is one of the most cost-effective voltage regulation methods for system planning. A joint planning and operation optimization algorithm was presented in \cite{45Karagiannopoulos2017,451Karagiannopoulos2016} to upgrade traditional expansion measures and consider the voltage regulation impact of DERs with smart inverters in the operation stage.  However, the authors in \cite{45Karagiannopoulos2017} implemented an offline optimal power flow to obtain the setpoints for the inverters to describe the inverter's real-time behavior in the operation stage. They did not implement the real-time operational characteristics of the smart inverters in the planning process, which is more accurate.

This paper proposes a novel planning strategy to optimally place a minimum number of PV smart inverters with Volt-VAr control among the existing PV systems while mitigating the possible voltage issues in the distribution systems. The problem is formulated as a two-stage stochastic mixed-integer programming optimization model considering the worst voltage violation scenarios. The first stage is to place the minimum number of PV smart inverters with Volt-VAr control. The second stage describes the Volt-VAr control impact on the system voltage profile while minimizing the PVs' active power curtailments. The placed PV smart inverters work in the VAr priority mode and follow their own pre-defined Q-V curve to autonomously control the local voltage without communicating with other devices. To validate the system voltage stability with the autonomous inverter control, a detailed dynamic model of the PV inverter is developed as a dynamic link library (DLL) in OpenDSS. 
The key contribution of this work can be summarized as follows:

\begin{enumerate}
\item A stochastic decision process is proposed to optimally place the PV smart inverters with Volt-VAr control considering the uncertainties of PV generation and load demand. 
	\item A set of analytical constraints are formulated in the second stage to model the impact of the Volt-VAr control with VAr priority on the voltage profile. The operational characteristics of the Q-V curve are defined as a piecewise-linear function and transformed into a block of variables and constraints to enforce the relationship between the reactive power output of the Volt-VAr controller and the local voltage magnitude. 
	
	\item A detailed dynamic model of the PV inverter with Volt-VAr control is developed as a DLL in OpenDSS to verify the optimization results and ensure system voltage stability.
	
	\item The optimization has been applied on an actual distribution feeder with instantaneous penetration levels as high as $200\%$ with significant overvoltage issues. The results show that a small subset of inverters upgraded with Volt-VAr control capability is sufficient to remove all voltage violations.

\end{enumerate}

This paper is organized as follows: Section II describes the problem to be solved in this paper. Section III presents the detailed mathematical formulation. Section IV provides the simulation results and the results verification. Section V summarizes the conclusion.

\section{Problem Statement} \label{PS}
\begin{figure}
	\centering
	\includegraphics[width=3.1in]{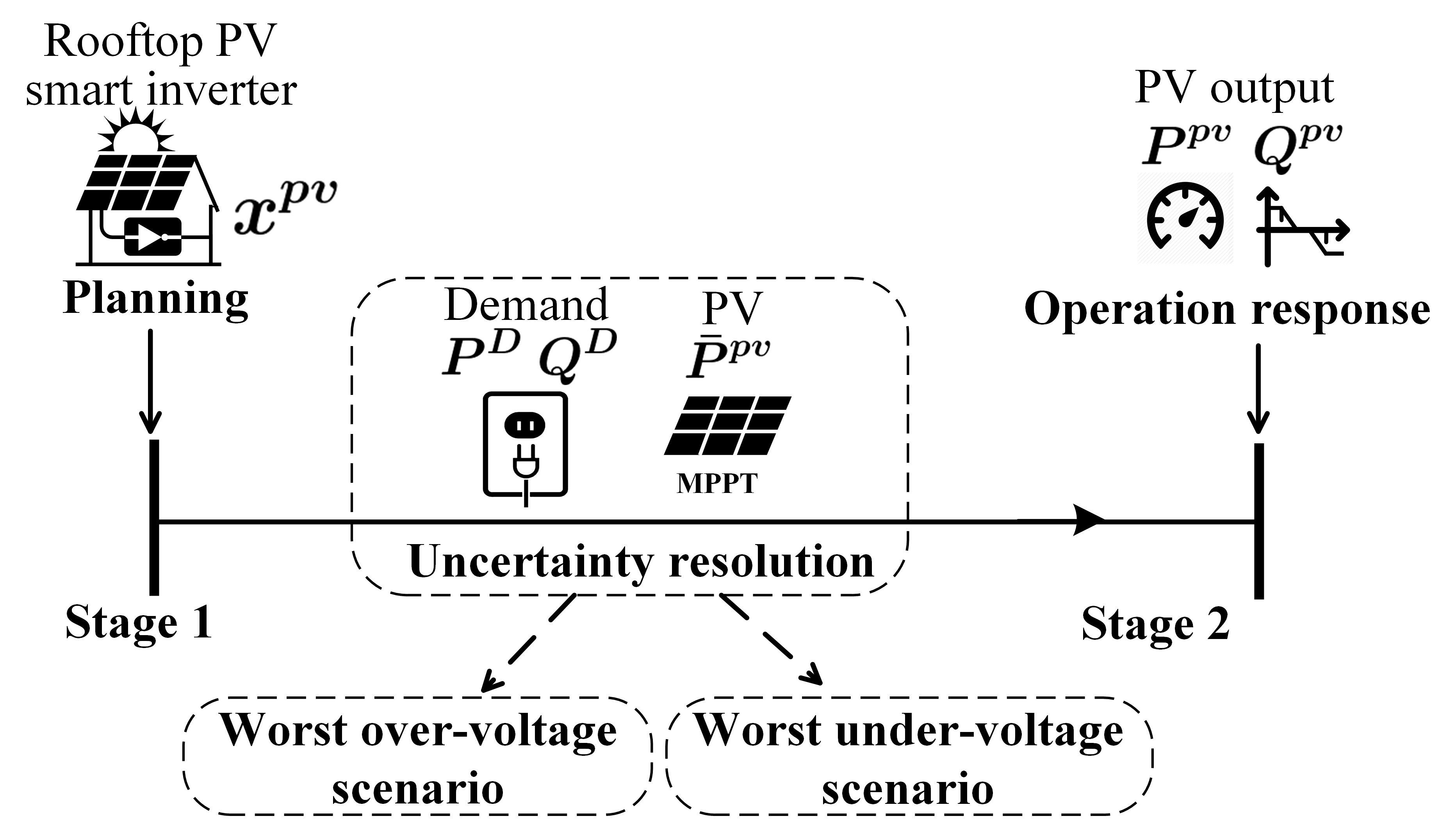}
	\caption{Decision process of placing PV smart inverter} 
        \vspace{-0.2in}
	\label{fig_decision}
\end{figure}
The increased penetration of PVs in distribution systems may lead to severe voltage violation problems or reactive power problems \cite{Sun2021}. PVs with smart inverters can control and optimize the local voltage by injecting or absorbing reactive power based on the local voltage. However, many of the PVs were not equipped with smart inverters. Retrofitting all or a large number of existing inverters to have Volt-VAr capability can be cost-prohibitive for the utilities. On the other hand, to ensure that the smart inverter can successfully control the distribution system voltage profile, especially during periods of peak solar output, the smart inverter may need to operate under VAr priority mode. It indicates that the reactive power support is prioritized, the active power output may need to be curtailed due to the lack of sufficient headroom in the inverter rating. However, curtailing the active power output of PVs negates the economic benefit to PV owners and other environmental benefits. An optimal placement strategy to place the minimum number of PV smart inverters with Volt-VAr control considering the uncertainties of PV output and load needs to be developed to solve these problems.  

As depicted in Fig. \ref{fig_decision},  the PV smart inverter placement problem is modeled as a two-stage stochastic decision process: (i) the planner makes smart inverter placement decisions for the rooftop PV in the first stage; (ii) the operational uncertainties are resolved in the worst-case voltage conditions, including (a) power demand and (b) the maximum power point of the PV output; the operator invokes recourse decisions (i.e., Volt-VAr control) to minimize voltage violation in the second stage. Here we use the historical data from a utility to construct the uncertainty, considering two worst voltage scenarios: The worst over-voltage scenario occurs under the maximum generation condition; the worst under-voltage scenario considers the maximum load condition.

\section{Mathematical Formulation}\label{MF}
	This section presents a two-stage stochastic mixed-integer linear program (SMILP) formulation to place the minimum number of PV smart inverters with Volt-VAr control to meet the voltage requirements and mitigate under/over-voltage conditions. The first stage minimizes the number of PV smart inverters. The second stage minimizes the squared norm of PVs' active power curtailment while maintaining the substation voltage at a certain level given the worst voltage scenario. Equation \eqref{Prep_Obj} presents the objective of the proposed SMILP problem, which minimizes the allocation cost of PV smart inverters and the expected operation cost of the second stage.
\small
\begin{equation}
	\begin{gathered}
		\min	~w_c\!\!\sum_{(g,p)\in \Omega_\mathcal{PV}}\!x^{pv}_{g,p} + \sum_{s\in \mathcal{S}}p_r(s)\phi(x^{pv},s)\label{Prep_Obj}
	\end{gathered}
\end{equation}
\normalsize
\begin{figure}
	\centering
	\includegraphics[width=2.8in]{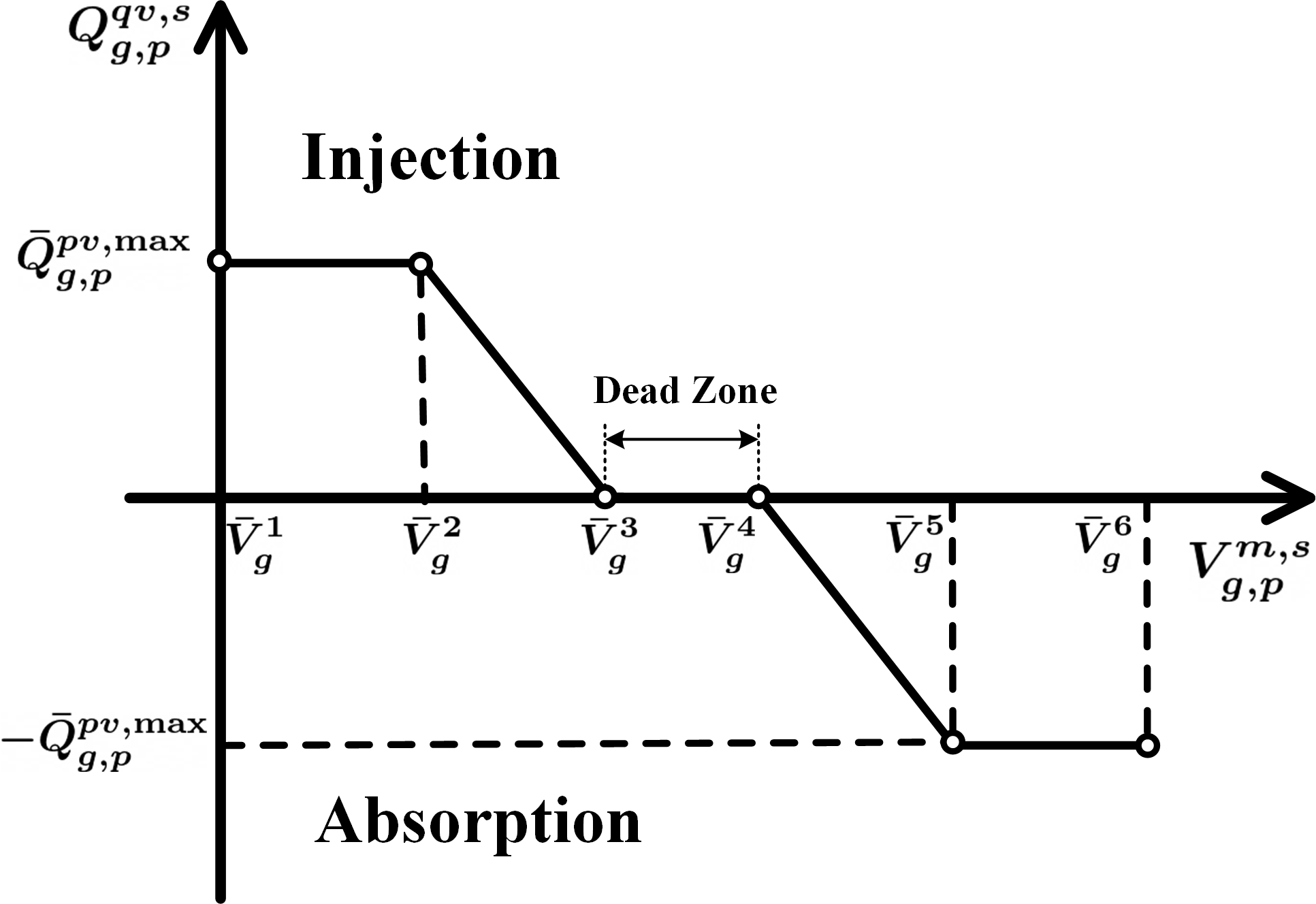}
	\caption{Q-V Curve of PV with Volt-VAr control} 
	\label{fig_qv}
        \vspace{-0.3cm}
\end{figure}
The second stage  models the unbalanced distribution system operation under worst case voltage scenarios. There are two modeling challenges. Firstly, if a PV smart inverter works in the Volt-VAr control mode with VAr priority, the generated reactive power should follow a specific Q-V curve as shown in Fig. \ref{fig_qv}. The inverter operates in different modes generating or absorbing reactive power to support local voltage. This Volt-VAr control function can be analytically achieved using the piece-wise linear Q-V curve constraint. At the same time, the squared norm of active power curtailment is minimized in the objective to encourage as many PV smart inverters as needed to participate in voltage violation mitigation through reactive power support.

Secondly, a multi-phase optimal power flow formulation that models the unbalanced distribution system accurately is needed to obtain the optimal location of PV smart inverters with Volt-VAr control based on actual system requirements. The simulation results in OpenDSS software at the same operation point are used to validate the solution from the optimization. This paper makes full use of the unbalanced distribution system linearized AC power flow formulation proposed in \cite{Zahra2022} to model all details of a distribution network. This linearized power flow model is based on the rectangular Current-Voltage (IV) formulation and uses the first-order approximation of the Taylor's series expansion to linearize the nonlinear product of current and voltage in the node power balance constraint. In order to construct the Q-V curve constraints for modeling the reactive power of PV smart inverters with Volt-VAr control, the voltage magnitude is needed, which can be obtained using a nonlinear function of real and imaginary parts of voltage in the IV formulation. We use a similar first-order approximation of the Taylor's series expansion to get the linear expression of the voltage magnitude.  

The detailed mathematical formulation of the second-stage problem is described as follows:

\subsubsection{Second-Stage Formulation}
\small
\begin{equation}
	\begin{gathered}
		\phi(x^{pv},s)=\min ~w_o^s\!\!\!\sum_{{(g,p)\in\Omega_\mathcal{PV}}}\!\!\!\left(P^{pv,s}_{g,p}-{\bar{P}}^{pv,s}_{g,p}\right)^2 \\  + w_v^s\!\!\!\sum_{(n,p)\in\Omega_{s\!u\!b}}\!\!\!\left(\left(V_{n,p}^{r,s}-{\bar{V}}_{n,p}^{r,s}\right)^2  +\left(V_{n,p}^{im,s}-{\bar{V}}_{n,p}^{im,s}\right)^2 \right) \label{second-stage obj}
	\end{gathered}
\end{equation}
\normalsize
The first part of the detailed second-stage formulation in \eqref{second-stage obj} minimizes the weighted least squares of the active power curtailment of PV with the smart Volt-VAr control. The second set of terms minimizes the weighted least squares of the feeder-head voltage difference between the optimization model and the base case without PV Volt-VAr control. 

\subsubsection{Smart inverter Volt-VAr Control Constraints}

\paragraph{PV Volt-VAr control enabling constraint}
\small
\begin{equation}
	\begin{gathered}
		x^{pv_{on}}_{g,p} \leqslant x^{pv}_{g,p}, \forall (g,p)\!\in{\Omega_\mathcal{PV}}_{vv}, s\in \mathcal{S} \label{x_pv_on}
	\end{gathered}
\end{equation}
\normalsize

Constraint \eqref{x_pv_on} indicates that only if a PV smart inverter is placed at the selected PV bus node $g$ phase $p$, the Volt-VAr control function can be enabled; otherwise, it will not be activated. Here ${\Omega_\mathcal{PV}}_{vv}$ is a pool of candidate PVs in the optimization that can be selected to install a PV smart inverter.

\paragraph{PV output disjunctive constraints}
\small
\begin{equation}
	\begin{gathered}
		\left\lceil\begin{matrix}\begin{matrix}x^{pv_{on},s}_{g,p}\\0\leqslant P^{pv,s}_{g,p}\leqslant {\bar{P}}^{pv,s}_{g,p}\\Q^{pv,s}_{g,p}=Q^{qv,s}_{g,p}\\\end{matrix}\\ {P^{pv,s}_{g,p}}^2+{Q^{pv,s}_{g,p}}^2\leqslant {(\bar{S}^{pv}_{g,p})}^2\\\end{matrix}\right\rceil \bigvee_{\substack{
				\forall (g,p)\in{\Omega_\mathcal{PV}}_{vv}\\ 
				s\in\mathcal{S}}} 	\left\lceil\begin{matrix}\begin{matrix}x^{pv_{o\!f\!f},s}_{g,p}\\ P^{pv,s}_{g,p}= {\bar{P}}^{pv,s}_{g,p}\\Q^{pv,s}_{g,p}=0\\\end{matrix}\\\end{matrix}\right\rceil \\
			x^{pv_{on},s}_{g,p}, x^{pv_{o\!f\!f},s}_{g,p}\!\!\in\!\left\{\text{True, False}\right\},\forall (g,p)\!\in\!{\Omega_\mathcal{PV}}_{vv},s\!\in\!\mathcal{S} \label{disjuction}
	\end{gathered}
\end{equation}
\normalsize

The disjunction constraint \eqref{disjuction} describes the logical relationship of the PV output with a binary decision, whether to enable Volt-VAr control function or not. Here $	x^{pv_{on},s}_{g,p}$ and $x^{pv_{o\!f\!f},s}_{g,p}$ are used as binary variables to select between different groups of  PV output constraints. These binary variables must satisfy the relationship $x^{pv_{on},s}_{g,p}\!+\!x^{pv_{o\!f\!f},s}_{g,p}\!=\!1$. If $x^{pv_{on},s}_{g,p}\!=\!1$,  it indicates that the PV at bus $g$ can perform Volt-VAr control: the active power output of the PV can be curtailed; the reactive power output of the PV needs to follow the Q-V curve as shown in Fig. \ref{fig_qv}, which varies with the voltage magnitude $V^m_{g,p}$; and the apparent power output of PV should be less than the rated apparent power value $\bar{S}^{pv}_{g,p}$.   If $x^{pv_{o\!f\!f},s}_{g,p}\!=\!1$, it indicates that the PV at bus $g$ does not perform Volt-VAr control:  the active power output of the PV is equal to the maximum power point for the PV, and its reactive power output is zero. 

\paragraph{Piecewise-linear Q-V curve constraint} \label{Q_V_Curve_constraints}  
The Q-V curve of Volt-VAr control shown in Fig. \ref{fig_qv} can be expressed as a continuous piecewise-linear function:
\small
\begin{equation}
	\begin{gathered}
		Q_{g,p}^{qv,s}=\left\{\begin{matrix}
			\bar{Q}_{g,p}^{pv,\max}& \ \text{if}~ \bar{V}^1_g \leqslant V^{m,s}_{g,p} \leqslant \bar{V}^2_g\\ 
			\frac{\bar{Q}_{g,p}^{pv,\max}}{\bar{V}_g^2-V_g^1}\left(\bar{V}_g^2 -V^{m,s}_{g,p}\right) & \text{if}~ \bar{V}^2_g \leqslant V^{m,s}_{g,p} \leqslant \bar{V}^3_g\\ 
			0 & \text{if}~ \bar{V}^3_g \leqslant V^{m,s}_{g,p} \leqslant \bar{V}^4_g\\ 
			\frac{\bar{Q}_{g,p}^{pv,\max}}{\bar{V}_g^3-\bar{V}_g^4}\left(V^{m,s}_{g,p}-\bar{V}_g^3\right) & \text{if}~ \bar{V}^4_g \leqslant V^{m,s}_{g,p} \leqslant \bar{V}^5_g\\ 
			-\bar{Q}_{g,p}^{pv,\max}& \text{if}~ \bar{V}^5_g \leqslant V^{m,s}_{g,p} \leqslant \bar{V}^6_g
		\end{matrix}\right. \label{eq_qv}
	\end{gathered}
\end{equation}
\normalsize

Here we use the disaggregated convex combination model to transform this function into a block of variables and constraints that enforces a piece-wise linear relationship between the voltage magnitude variable (i.e., $V_{g,p}^{m,s}$)  and the PV reactive power output variable (i.e., $Q^{qv,s}_{g,p}$). Let ${\bar{Q}}_g^j\in\left\{\bar{Q}_{g,p}^{pv,\max},\bar{Q}_{g,p}^{pv,\max},0,0,-\bar{Q}_{g,p}^{pv,\max},-\bar{Q}_{g,p}^{pv,\max}\right\}$ represents the corresponding reactive output at the voltage magnitude breakpoints of the Q-V curve. The reformulated constraints of \eqref{eq_qv} can be written as:
\small
\begin{equation}
	\begin{gathered}
		V^{m,s}_{g,p}\!\!=\!\!\sum_{j\in\mathcal{J}} \left(\bar{V}^j_g\lambda_{g,p}^{j,j,s}+\bar{V}^{j+1}_g\lambda_{g,p}^{j,j+1,s}\right) ,\forall (g,p) \in \Omega_\mathcal{PV},s\in \mathcal{S}
	\end{gathered}
\end{equation}
\begin{equation}
	\begin{gathered}
		Q_{g,p}^{qv,s}\!\!=\!\!\sum_{j\in\mathcal{J}} \left(\bar{Q}^j_g\lambda_{g,p}^{j,j,s}+\bar{Q}^{j+1}_g\lambda_{g,p}^{j,j+1,s}\right) ,\forall (g,p) \in \Omega_\mathcal{PV},s\in \mathcal{S}
	\end{gathered}
\end{equation}
\begin{equation}
	\begin{gathered}
		\delta_{g,p}^{j,s} =  \lambda_{g,p}^{j,j,s}+\lambda_{g,p}^{j,j+1,s} ,\forall (g,p) \in \Omega_\mathcal{PV}, j \in \mathcal{J},s\in \mathcal{S}
	\end{gathered}
\end{equation}
\begin{equation}
		\sum_{j\in\mathcal{J}}\delta_{g,p}^{j,s} =  1,\forall (g,p) \in \Omega_\mathcal{PV}, s\in \mathcal{S}
\end{equation}
\begin{equation}
	\delta_{g,p}^{j,s} \in \left\{0,1\right\},\forall (g,p) \in \Omega_\mathcal{PV},  j \in \mathcal{J}, s\in \mathcal{S}
\end{equation}
\begin{equation}
	\lambda_{g,p}^{j,k,s} \geqslant 0,\forall (g,p) \in \Omega_\mathcal{PV},  (j,k) \in \mathcal{K}, s\in \mathcal{S}
\end{equation}
\normalsize
where $\mathcal{J}$ is the set of line segments of the Q-V curve, and $|\mathcal{J}|$ is equal to 5. $\delta_{g,p}^{j,s}$ represents the binary variable selection of line segment $j$. $\lambda_g^{j,k,s}, \forall (j,k)\in \mathcal{K}$ is defined as a weight variable to perform the interpolation on the selected segment $j$, where $k$ represents the two end points belonging to line segment $j$ and $\mathcal{K}$ is the mapping set of the connecting line segment and its corresponding end-point.
From the Volt-VAr control setting aspect, $\bar{V}_g^2, \bar{V}_g^3,\bar{V}_g^4$, and $\bar{V}_g^5$ are the setpoints for the Q-V control curve, which can be adjusted in a specific allowable range. The maximum reactive power of the PV with the Volt-VAr control in \eqref{eq_qv} can be assumed to be the following:
\small
\begin{equation}
	\bar{Q}_{g,p}^{pv,max}\!=\!\sqrt{\left(\frac{1}{{P\!F_{\min}}^2}-1\right)}\bar{P}_{g,p}^{pv,max},\forall (g,p) \in\Omega_{\mathcal{PV}_{v\!v}} \label{Qmax}
\end{equation}
\normalsize
where $PF_{\min}$ represents the minimum power factor that the inverter is capable of operating at rated active power and $P_{g,p}^{pv,max}$ is the maximum active power output of the g-th PV source among all the time-series values given in the data. Based on \eqref{Qmax}, the PV's apparent power rating in \eqref{disjuction} can be assumed to be:
\small
\begin{equation}
	\bar{S}^{pv}_{g,p}=\sqrt{\bar{Q}^{pv,\max}_{g,p}{}^2+\bar{P}^{pv,\max}_{g,p}{}^2},\forall (g,p)\in\Omega_{\mathcal{PV}_{v\!v}} \label{Spv}
\end{equation}
\normalsize

\subsubsection{Linearized IV-based AC Power Flow Constraints}

The objective of mitigating voltage violations under the worst voltage scenarios with the minimum number of PV smart inverters with Volt-VAr control makes the placement decisions very sensitive to the voltage change. As a result, it is necessary to accurately model the three-phase unbalanced distribution system power flow considering the impact of lines' mutual impedances and shunt admittances on the bus voltage. Compared with the widely-used linearized DistFlow model, which approximates the unbalanced voltage as balanced  \cite{Robbins2016,zhang2019}, the IV-based AC power flow model proposed in \cite{Zahra2022} has been validated to be more accurate to capture the operation characteristics of the unbalanced distribution systems with a high penetration level of PVs. 
\paragraph{Line flow constraints} 
\small
\begin{equation}
	\begin{gathered}
	V^{r,s}_{\ell_o,p}-V^{r,s}_{\ell_e,p} =	\sum_{m\in \varphi(\ell)} R_{\ell,p,m}\bigg(I^{r,s}_{\ell,m}+\sum_{k\in\varphi(\ell)}y_{\ell,m,k}V^{im,s}_{\ell_o,k}\bigg)\\\!\!-\!\!\sum_{m\in \varphi(\ell)} \!\!\!X_{\ell,p,m}\!\bigg(I^{im,s}_{\ell,m}\!\!-\!\!\!\sum_{k\in\varphi(\ell)}y_{\ell,m,k}V^{r,s}_{\ell_o,k}\bigg),\forall (\ell,p) \in \Omega_\mathcal{L}, s\in \mathcal{S}
	\end{gathered} \label{lineflow_p}
\end{equation}
\begin{equation}
	\begin{gathered}
		V^{im,s}_{\ell_o,p}-V^{im,s}_{\ell_e,p} =	\sum_{m\in \varphi(\ell)} R_{\ell,p,m}\bigg(I^{im,s}_{\ell,m}+\sum_{k\in\varphi(\ell)}y_{\ell,m,k}V^{r,s}_{\ell_o,k}\bigg)\\\!\!-\!\!\sum_{m\in \varphi(\ell)} \!\!\!X_{\ell,p,m}\!\!\bigg(I^{r,s}_{\ell,m}\!\!-\!\!\!\sum_{k\in\varphi(\ell)}y_{\ell,m,k}V^{im,s}_{\ell_o,k}\bigg),\forall (\ell,p) \in \Omega_\mathcal{L}, s\in \mathcal{S}
	\end{gathered} \label{lineflow_q}
\end{equation}
\normalsize
Constraints \eqref{lineflow_p}-\eqref{lineflow_q} describe the relationship of line current and corresponding bus voltage difference over a distribution line connecting bus $\ell_o$ to bus $\ell_e$ considering the impact of the line's self and mutual impedances and admittances. 
\paragraph{Current injection constraints}
\small
\begin{equation}
	I_{i,p}^{in,r,s}\!=\!\sum_{(\ell,p)\epsilon \mathcal{L}_O\left(i\right)}\!\!\!\!\!\!I_{\ell,p}^{r,s}\!-\!\sum_{(\ell,p) \epsilon\mathcal{L}_E\left(i\right)}\!\!\!I_{\ell,p}^{r,s}, \forall (i,p) \in \Omega_\mathcal{B}, s\in \mathcal{S}
\end{equation}
\begin{equation}
	I_{i,p}^{in,im,s}\!\!=\!\!\!\sum_{(\ell,p)\epsilon \mathcal{L}_O\left(i\right)}\!\!\!\!\!\!I_{\ell,p}^{im,s}\!-\!\!\!\!\sum_{(\ell,p) \epsilon\mathcal{L}_E\left(i\right)}\!\!\!\!\!\!I_{\ell,p}^{im,s}, \forall (i,p) \in \Omega_\mathcal{B}, s\!\in\!\mathcal{S}
\end{equation}
\normalsize
The injected current at each node and scenario is obtained using (16)-(17).

\begin{figure*}[!ht]
	\centering
 \vspace{-0.1in}
	\includegraphics[width=1.5\columnwidth]{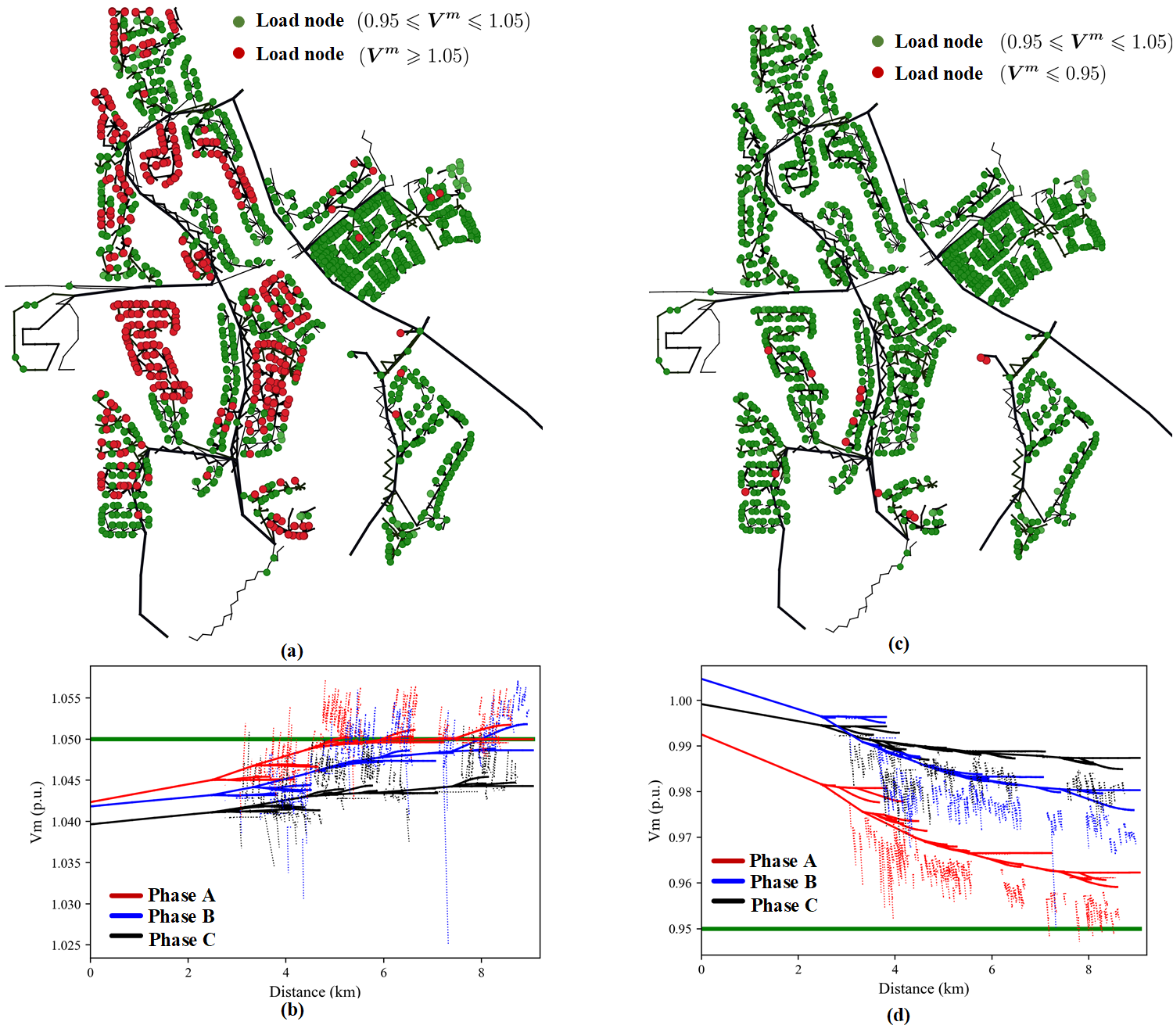}
  \vspace{-0.2in}
	\caption{(a) The voltage map in the over-voltage scenario; (b) The voltage profile in the over-voltage scenario; (c) The voltage map in the  under-voltage scenario; (d) The voltage profile in the under-voltage scenario} 
	\label{fig_overvoltage_wo_vv}
 \vspace{-0.2in}
\end{figure*}
\paragraph{Linearized power balance constraints}
In the rectangular IV formulation, the power balance constraints contain nonlinear elements due to the product of voltage and injected current. To linearize the power balance constraints, an iterative first-order approximation of the Taylor's series expansion developed in \cite{Zahra2022} is used and the linearized power balance constraints around an operating point for each phase (i.e., ${\hat{V}}_{i,p}^{r,s},{\hat{V}}_{i,p}^{im,s},\hat{I}_{i,p}^{in,r,s}$, and $\hat{I}_{i,p}^{in,im,s}$) can be expressed as \eqref{power_balance_p} and \eqref{power_balance_q}:
\small
\begin{equation}
	\begin{gathered}
			 \sum_{{\substack{\forall (n,p) \in \Omega_{sub}\\n=i}}}\!\!\!\!\!P^{G,s}_{n,p}\!\!+\!\!\!\sum_{{\substack{\forall (g,p) \in \Omega_{\mathcal{PV}_{vv}}\\g=i}}}\!\!\!\!P^{pv,s}_{g,p}\!\!\!+\!\!\!\!\sum_{{\substack{\forall (g,p) \in \Omega_{\mathcal{PV}_{wo}}\\g=i}}}\!\!\!\!{\bar{P}}^{pv,s}_{g,p}\!\!\!-\!\!\!\!\!\sum_{{\substack{\forall (d,p) \in \Omega_\mathcal{D}\\d=i}}}\!\!\!\!\!\!\!P^{D,s}_{d,p}\\-\sum_{{\substack{\forall (m,p) \in \Omega_{TR}\\m=i}}}\!\!\!\!\!\!P^{Tr,s}_{m,p}={\hat{V}}_{i,p}^{r,s}I_{i,p}^{in,r,s}+{\hat{I}}_{i,p}^{in,r,s}V_{i,p}^{r,s}
		+{\hat{V}}_{i,p}^{im,s}I_{i,p}^{in,im,s}\\+
			{\hat{I}}_{i,p}^{in,im,s}V_{i,p}^{im,s}\!\!\!-\!\!{\hat{V}}_{i,p}^{r,s}{\hat{I}}_{i,p}^{in,r,s}-\!{\hat{V}}_{i,p}^{im,s}{\hat{I}}_{i,p}^{in,im,s},\forall (i,p)\!\in\!\Omega_\mathcal{B},s\!\in\!\mathcal{S} \label{power_balance_p}
	\end{gathered}  
\end{equation}
\begin{equation}
	\begin{gathered}
		\sum_{{\substack{\forall (n,p) \in \Omega_{sub}\\n=i}}}\!\!\!\!\!Q^{G,s}_{n,p}\!\!+\!\!\sum_{{\substack{\forall (g,p) \in \Omega_{\mathcal{PV}_{v\!v}}\\g=i}}} \!\!\!\!\!\!\!Q^{pv,s}_{g,p}-\!\!\!\sum_{{\substack{\forall (d,p) \in \Omega_\mathcal{D}\\d=i}}} \!\!\!\!Q^{D,s}_{d,p}\!\!=\!\!\!{\hat{V}}_{i,p}^{im,s}I_{i,p}^{in,r,s}\\
		-{\hat{V}}_{i,p}^{r,s}I_{i,p}^{in,im,s}+{\hat{I}}_{i,p}^{in,r,s}V_{i,p}^{im,s}-
		{\hat{I}}_{i,p}^{in,im,s}V_{i,p}^{r,s}-{\hat{V}}_{i,p}^{im,s}{\hat{I}}_{i,p}^{in,r,s}\\-{\hat{V}}_{i,p}^{r,s}{\hat{I}}_{i,p}^{in,im,s},\forall (i,p)\in\Omega_\mathcal{B},s\in \mathcal{S} \label{power_balance_q}
	\end{gathered}  
\end{equation}
\normalsize
Two PV operating points are considered in the proposed SMILP problem (i) without smart inverter and (ii) with smart inverter. Since the objective of this placement problem is to minimize the number of PV smart inverters to mitigate the voltage violations in the worst-case scenarios, the PV operating condition without placing smart inverters here is considered to be the base case. In constraints \eqref{power_balance_p}-\eqref{power_balance_q}, $({\hat{V}}_{i,p}^{r,s},{\hat{V}}_{i,p}^{im,s},\hat{I}_{i,p}^{in,r,s},\hat{I}_{i,p}^{in,im,s})$ are the real and imaginary parts of bus voltage and injected current at the operating point without smart inverter, which are used as the first-order approximation parameters of Taylor's series expansion and can be updated iteratively for getting the voltage and injected current at the operating point with smart inverter. Note in constraint \eqref{power_balance_p}, the active power output of the residential PVs that are not in the PV smart inverter candidate pool is assumed to follow the respective MPP value.

\paragraph{Voltage magnitude constraints}
 The voltage magnitude can be expressed by a nonlinear function of the real part and the imaginary part of voltage in \eqref{eq_Vm2}.
 \small
\begin{equation}
	V^{m,s}_{i,p}=\!\!\!\sqrt{{V^{r,s}_{i,p}}{}^2+{V^{im,s}_{i,p}}{}^2}, \forall (i,p)\in\Omega_\mathcal{B},s\in\mathcal{S} \label{eq_Vm2}
\end{equation}
\normalsize

The linear approximation of \eqref{eq_Vm2} can be reformulated by the similar first-order Taylor-series expansion method:
\small
\begin{equation}
	\begin{gathered}
			V^{m,s}_{i,p}\!\!\!=\!\!\!\sqrt{{\hat{V}^{r,s}_{i,p}}{}^2\!\!+\!\!{\hat{V}^{im,s}_{i,p}}{}^2} + \frac{\partial V^{m,s}_{i,p} }{\partial V^{r,s}_{i,p}}\bigg|_{\hat{V}^{r,s}_{i,p}} (V^{r,s}_{i,p}- \hat{V}^{r,s}_{i,p}) \\+\frac{\partial V^{m,s}_{i,p} }{\partial V^{im,s}_{i,p}}\bigg|_{\hat{V}^{im,s}_{i,p}} (V^{im,s}_{i,p}\!-\!\hat{V}^{im,s}_{i,p})\\
			=\frac{\hat{V}^{r,s}_{i,p}{V}^{r,s}_{i,p}}{\sqrt{{\hat{V}^{r,s}_{i,p}}{}^2\!\!+\!\!{\hat{V}^{im,s}_{i,p}}{}^2}}+			\frac{\hat{V}^{im,s}_{i,p}{V}^{im,s}_{i,p}}{\sqrt{{\hat{V}^{r,s}_{i,p}}{}^2\!\!+\!\!{\hat{V}^{im,s}_{i,p}}{}^2}}
	\end{gathered}
\end{equation}
\normalsize
where $(\hat{V}^{r,s}_{i,p},\hat{V}^{im,s}_{i,p})$ represent the operating point without smart inverter for the first iteration of Taylor's series expansion at bus $i$, phase $p$, and scenario $s$. 

Constraint \eqref{Vm} bounds the voltage magnitudes in the normal operating range. 
\small
\begin{equation}
\begin{gathered}
	V^{\min} \leqslant V^{m,s}_{i,p} \leqslant V^{\max}, \forall (i,p) \in \Omega_\mathcal{B},s\in \mathcal{S} \label{Vm}
\end{gathered}
\end{equation}
\normalsize

\section{Simulation Results and Validation} 
This section presents the optimal results of the proposed PV smart inverter placement problem corresponding to an actual 12.47 kV, 9 km-long Arizona utility feeder that serves residential customers. This feeder has 7864 buses, 1790 primary sections, 5782 secondary sections, 371
distribution transformers, 1737 loads, and 766 residential rooftop PV units. The detailed information of this feeder can be found in \cite{MontanoMartinez2021}. 
 \vspace{-0.1in}
\subsection{Scenario Generation and Solution Algorithm} 
As mentioned in Section \ref{PS}, two days corresponding to the actual historical feeder data - the maximum generation condition on  03/15/2019 (high PV) and load peak on 07/15/2019 (high load and relatively low PV) were chosen for constructing the worst over and under voltage scenarios. In the worst over-voltage scenario, the power generated by the installed residential rooftop PV is $3.6 \text{MW}$, which corresponds to a penetration level of $225\%$ ($3.6 \text{MW}/1.6 \text{MW}$) compared to the feeder's corresponding total gross load. Fig. \ref{fig_overvoltage_wo_vv} (a)(b) present the load node locations and the voltage profile with the over-voltage issue, and the total number of over-voltage load nodes is 439. For the worst under-voltage scenario, Fig. \ref{fig_overvoltage_wo_vv} (c)(d) present the load node locations and the voltage profile, and the total number of under-voltage load nodes is $14$. 

\begin{figure}[!ht]
	\centering
	\includegraphics[width=0.43\textwidth]{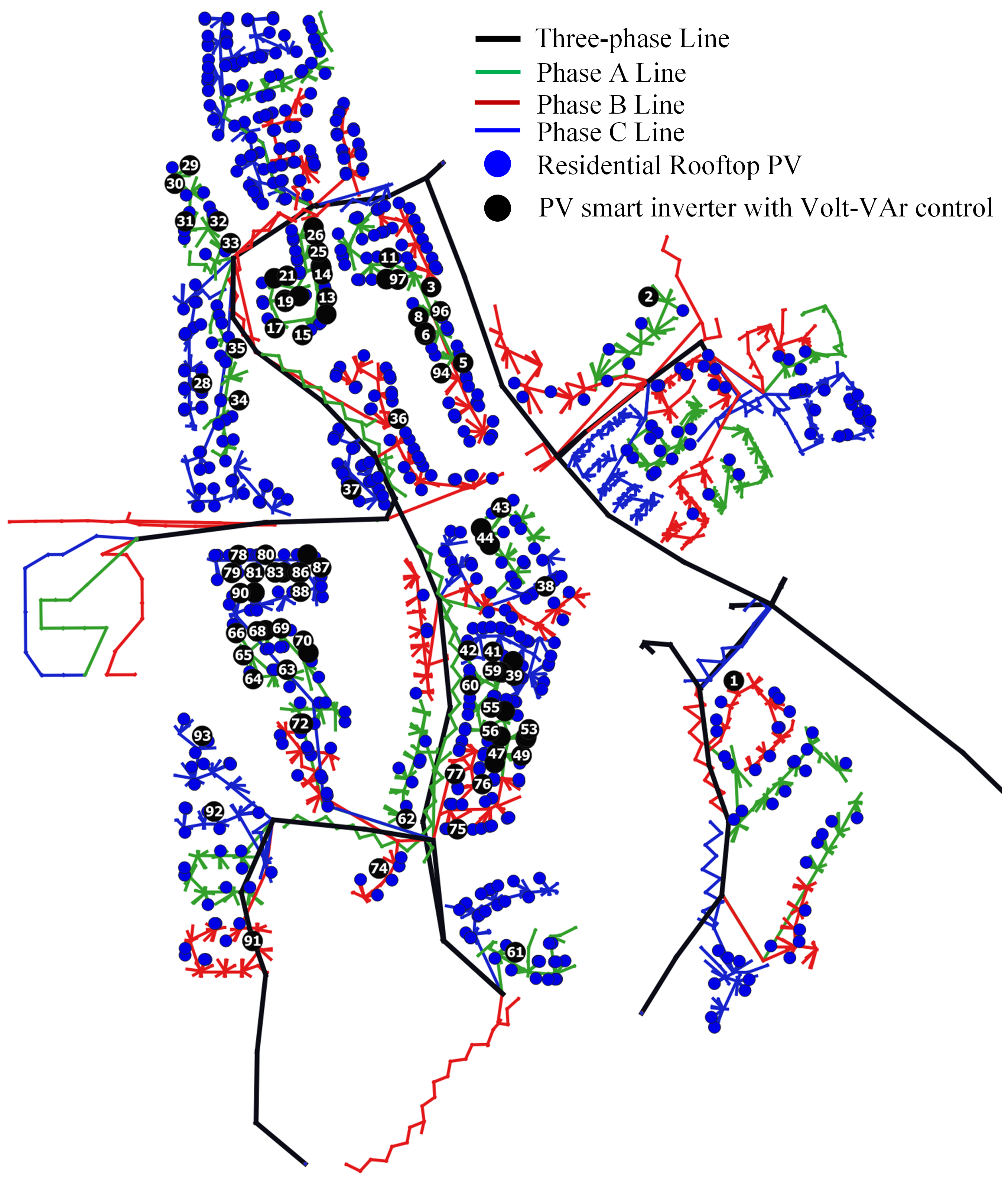}
  \vspace{-0.2in}
	\caption{The optimal locations of PV smart inverters with Volt-VAr control} 
	\label{fig_pv_location}
 \vspace{-0.1in}
\end{figure}

As only two voltage scenarios are considered, the extensive form of the proposed two-stage SMIP model for the optimal placement of PV smart inverters is directly solved using the PySP package in Pyomo (version 5.7.3) with Gurobi 9.03 mixed-integer solver. The simulations were performed on a computer with a 3.6 GHz 8-Core Intel i9-9900K CPU and 36 GB of RAM. 
\begin{figure}
	\centering
 \vspace{-0.1in}
	\includegraphics[width=0.4\textwidth]{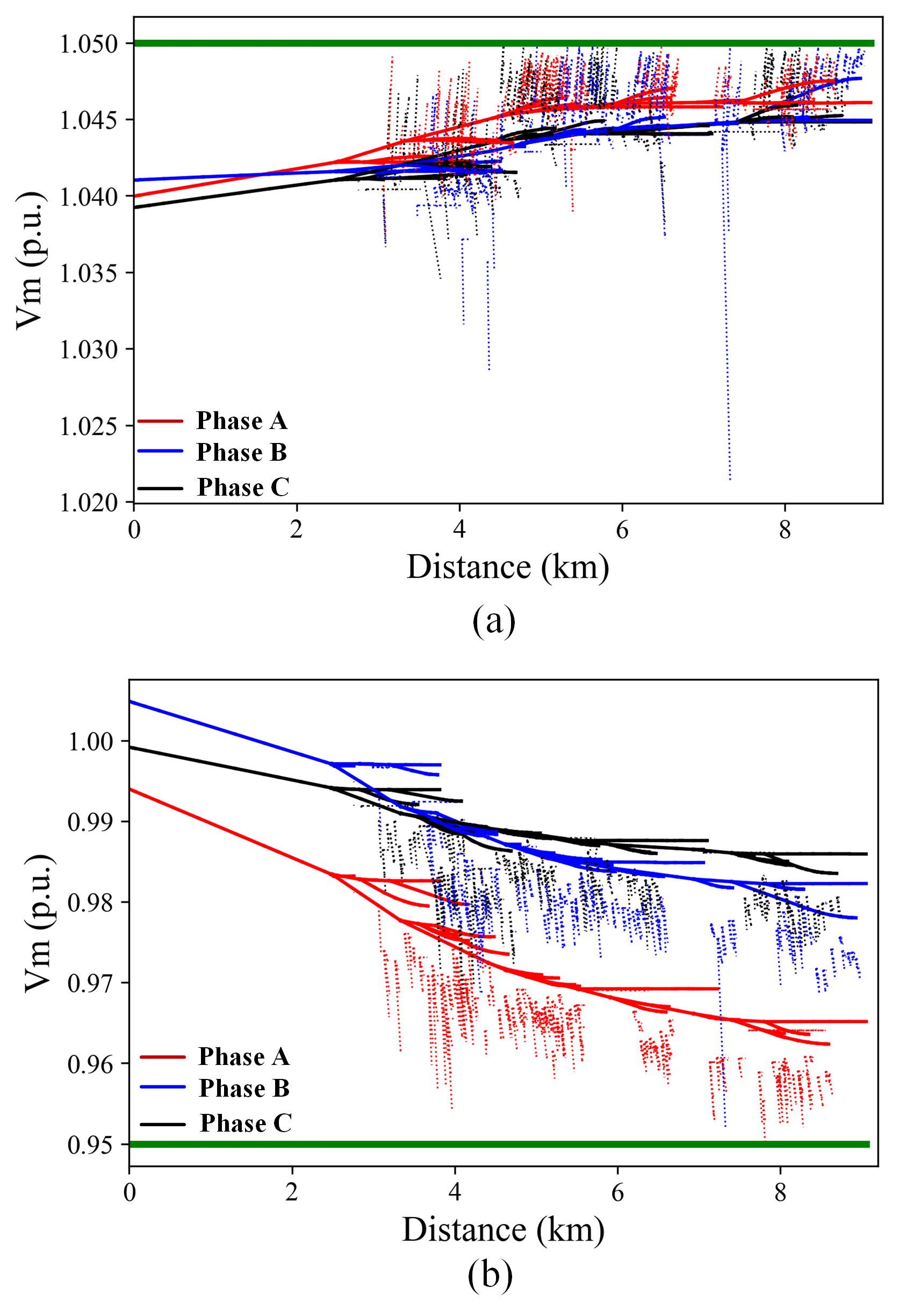}
	\caption{Voltage profile after enabling the optimally placed  Volt-VAr controller in (a) over-voltage scenario and (b) under-voltage scenario} 
	\label{fig_voltage_vv}
  \vspace{-0.1in}
\end{figure}

\subsection{The Operation Results Comparison of the SMIP Optimization with OpenDSS}
The six parameters of the Q-V curves of Volt-VAr control are set as $\bar{V}_g^1\!=\!0.0,\bar{V}_g^2\!=\!0.94,\bar{V}_g^3\!=\!0.98,\bar{V}_g^4\!=\!1.02,\bar{V}_g^5\!=\!1.06,\bar{V}_g^6\!=\!1.1, \forall g \!\in\!\Omega_{\mathcal{PV}_{\!v\!v}}$. The power factor range of PV smart inverter is $[-0.8,0.8]$. The optimal location of the PV smart inverters is shown in Fig. \ref{fig_pv_location} and the total number is 99. There are 8 PV smart inverters in Phase A, 69 in Phase B, and 22 in Phase C. Those selected PV smart inverters with Volt-VAr control are implemented in the OpenDSS to validate the effectiveness of the proposed SMIP model. By comparing Fig. \ref{fig_overvoltage_wo_vv} (a) and Fig. \ref{fig_pv_location}, it can be seen that the optimization typically favors the locations with overvoltages initially for the placement of smart inverters. The voltage profiles after enabling the optimally selected Volt-VAr control in the OpenDSS simulation under the two worst voltage scenarios are shown in Fig. \ref{fig_voltage_vv}. It can be seen that the optimally placed PV smart inverters with Volt-VAr control can successfully mitigate the voltage violation issues for both over-voltage and under-voltage scenarios.  
\begin{figure}
	\centering
	\includegraphics[width=0.40\textwidth]{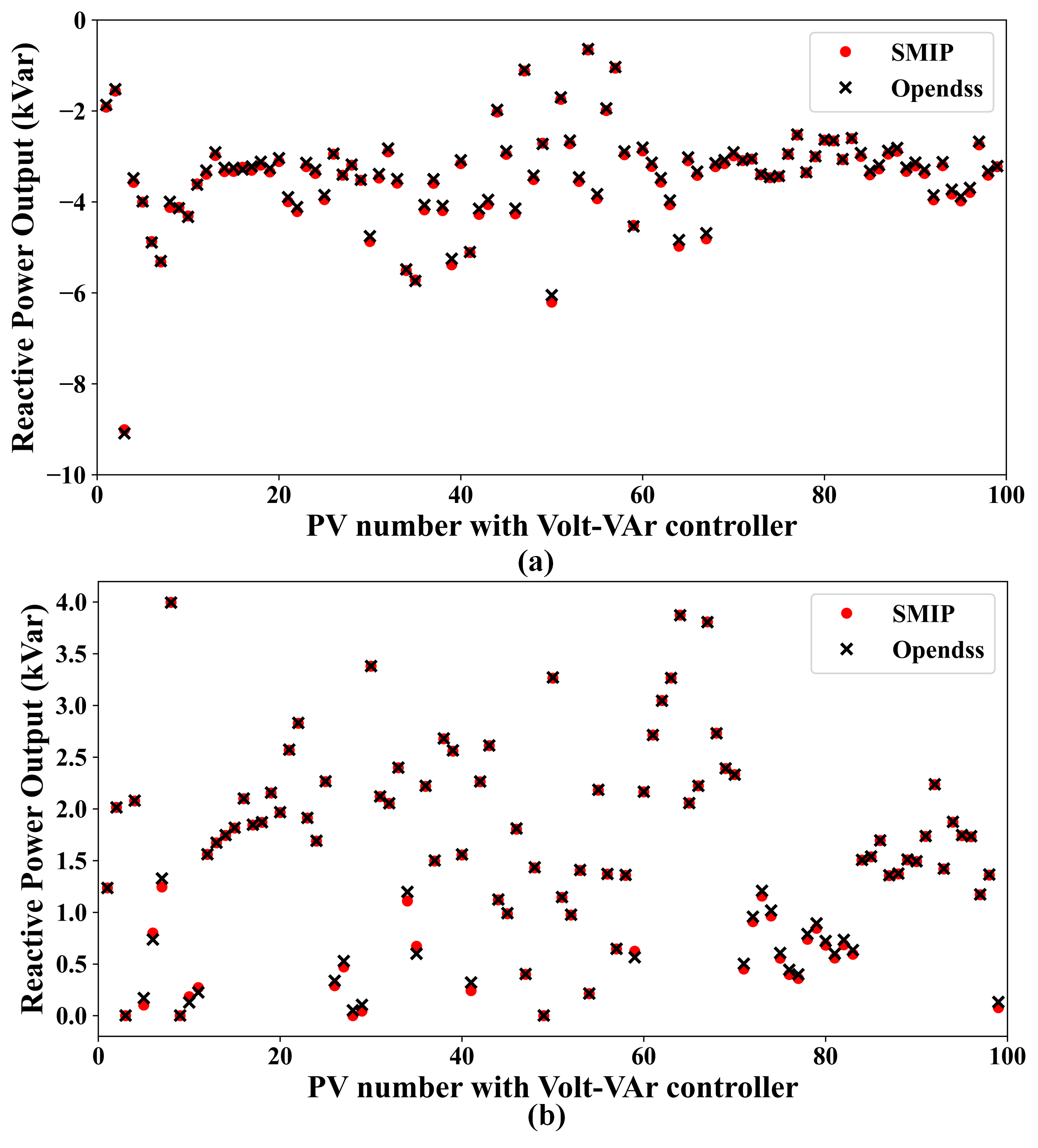} 
        \vspace{-0.2in}
	\caption{Reactive power output comparison for (a) over-voltage scenario and (b) under-voltage scenario}
	\label{fig_Q_compare}
  \vspace{-0.1in}
\end{figure}

\begin{figure}
	\centering
	\includegraphics[width=0.5\textwidth]{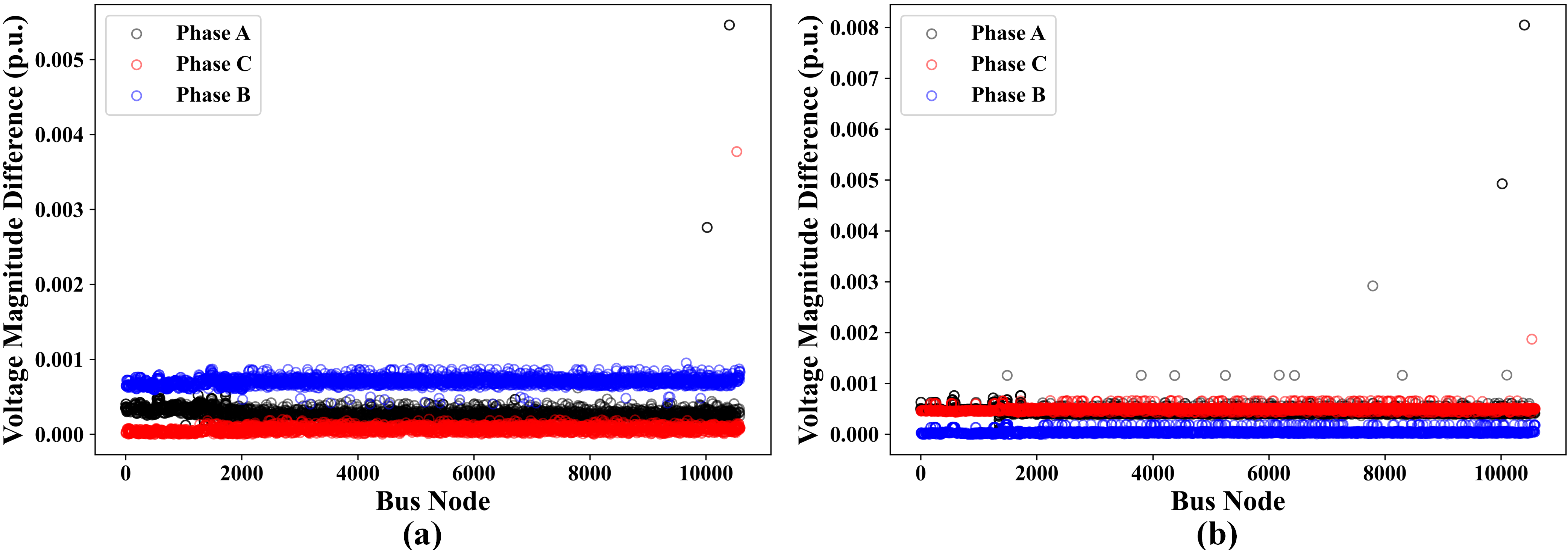}
	\vspace{-0.2in}
	\caption{The difference in voltage magnitudes obtained from SMIP model and OpenDSS at each bus for (a) over-voltage scenario and (b) under-voltage scenario}
	\label{fig_v_compare}
	\vspace{-0.1in}
\end{figure}

To validate the effectiveness of the proposed optimization model, we compare the power flow results in the optimization model with the OpenDSS simulation results after enabling Volt-VAr control in the selected PV smart inverters.  Figures \ref{fig_Q_compare} (a) and (b) present the PV smart inverters' reactive power comparison between the SMIP model and OpenDSS under both scenarios. The average squared difference of the PV smart inverter's reactive power output between the SMIP model and OpenDSS is  $0.548\%$ and $0.089\%$ respectively for the over-and under-voltage scenarios, which are relatively small values. The difference between the active power output of the smart inverters in the SMIP model and the maximum power point of PV generators in both scenarios is zero. It indicates that the optimally placed PV smart inverter with Volt-VAr control can guarantee the customers' economic benefit of maximizing their PV units' active power output under the worst voltage scenarios. At the same time, we compare the voltage magnitude difference of each bus node in both the SMIP model and OpenDSS simulation as shown in Fig. \ref{fig_v_compare}. These results validate the effectiveness of the proposed SMIP model in modeling the unbalanced distribution system operation considering the impact of PV smart inverter with Volt-VAr control on the system voltage profile.  

\begin{figure}
	\centering
	\includegraphics[width=0.4\textwidth]{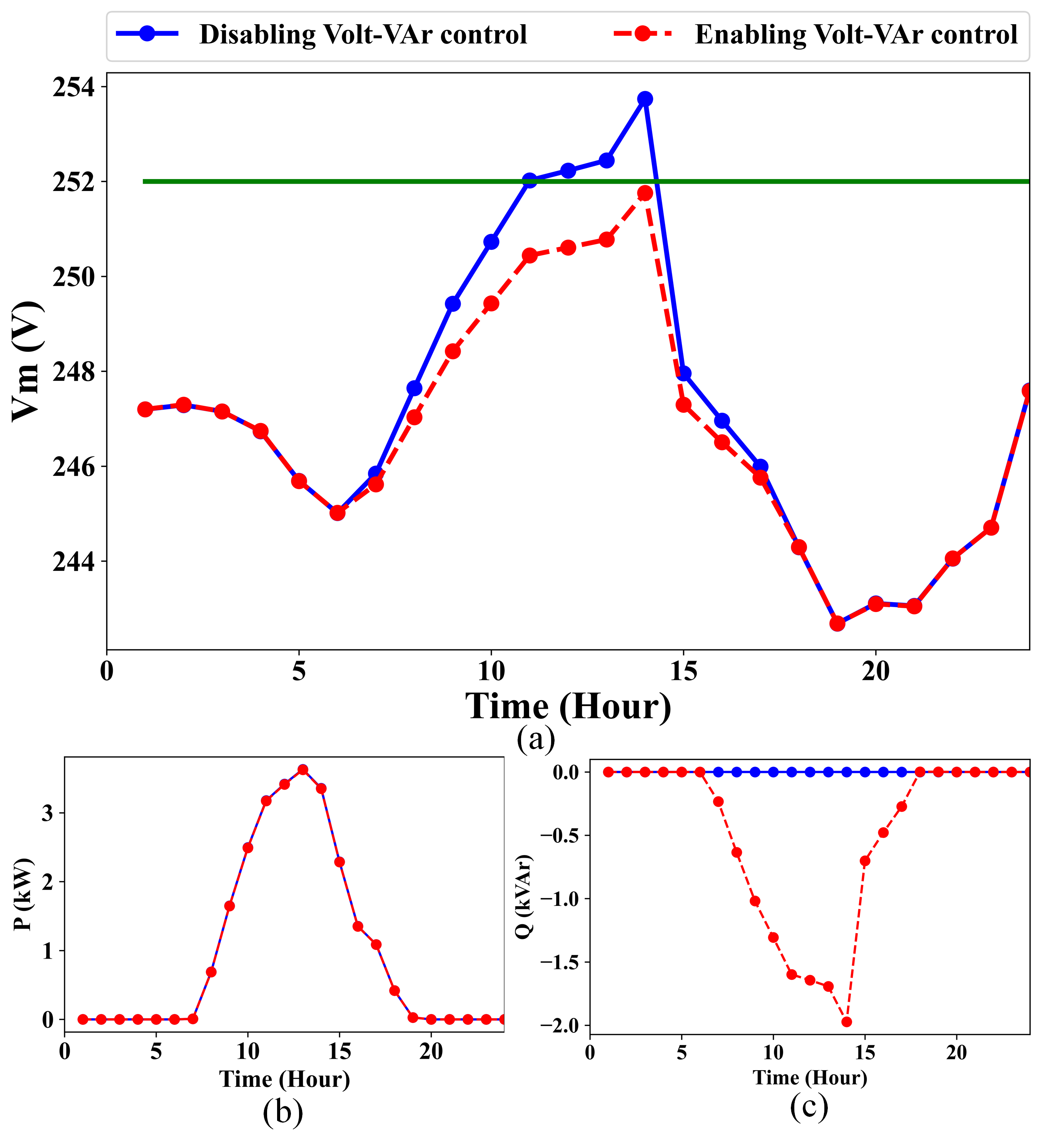}\vspace{-0.1in}
	\caption{24-hour time-series voltage magnitude, active power and reactive power output comparison for the 45th PV smart inverter disabling and enabling Volt-VAr control} 
 \vspace{-0.1in}
	\label{fig_V_24}
\end{figure}

As only the worst voltage violation scenarios are considered, it is necessary to validate whether the optimal placement decisions work for other time instants. Two 24-hour time-series power flow studies disabling and enabling Volt-VAr control at the selected $99$ PV smart inverters are conducted in the OpenDSS. Fig. \ref{fig_V_24} shows the hourly time-series voltage magnitude, active power output, and reactive power output comparison for the bus node with the 44th placed PV smart inverter. This specific PV bus has the maximum voltage violation in the over-voltage scenario. It is found that this bus node, when Volt-VAr control is disabled, violates the normal operation voltage limit from $t=11\text{h}$ and reaches its maximum voltage magnitude at $t=14 \text{h}$, which causes the worst over-voltage problem. With Volt-VAr enabled, there is no voltage violation issue during the entire 24-hour operation as the reactive power absorption of the PV smart inverters helps to reduce the voltage.

\subsection{The Dynamic Voltage Stability Validation}
%
\begin{figure}
	\centering
	\includegraphics[width=0.45\textwidth]{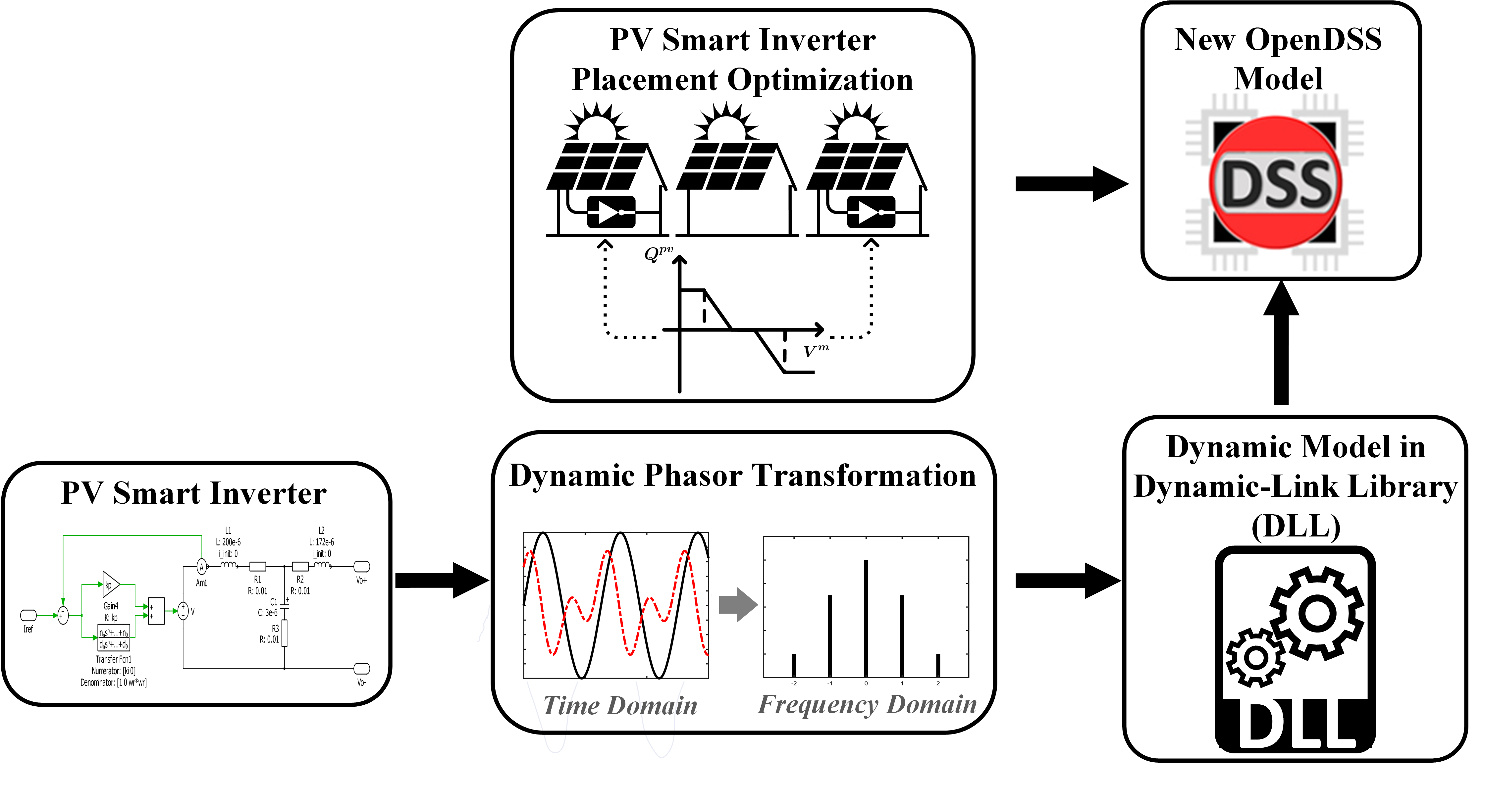}
  \vspace{-0.2in}
	\caption{The flow chart of optimization result validation in OpenDSS} 
	\label{validationflowchart} \vspace{-0.1in}
\end{figure}

To validate the voltage stability of the Volt-VAr control operation, a detailed dynamic model of the PV smart inverter with Volt-VAr control capability is developed based on previous work \cite{Yu2019}. Fig. \ref{validationflowchart} presents the flow chart of the entire validation process, which can be described as follows: (1) model the dynamics of the PV smart inverter with Volt-VAr control in the real-valued time domain; (2) adopt the dynamic phasor transformation to transform the real-valued inverter model to a phasor-based model implemented in a DLL; (3) test the system stability in both static and dynamic simulations to validate the effectiveness of the optimization results. 

\begin{figure}
	\centering
	\includegraphics[width=2.5in]{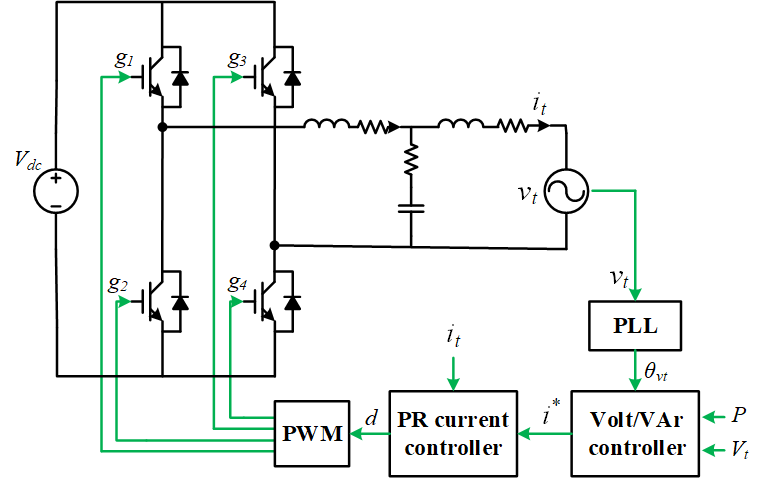}
	\caption{Block diagram of the PV inverter} 
	\label{fig_inv}\vspace{-0.2cm}
\end{figure}

Fig. \ref{fig_inv} shows the simplified block diagram of the inverter power stage used for validation, which is numerically implemented in the DLL. The PLL is used to extract the reference angle $\theta_{vt}$ from the terminal voltage $v_t$. The Proportional-Resonant (PR) current controller can force the grid-side inductor current, that is, the terminal current $i_t$ of the inverter, to follow the current reference $i^*$ provided by the Volt-VAr control block. The inverter's active and reactive power output can be controlled to conduct various grid support functions by adjusting active current and the reactive current separately. 

\begin{figure}
	\centering
	\includegraphics[width=0.45\textwidth]{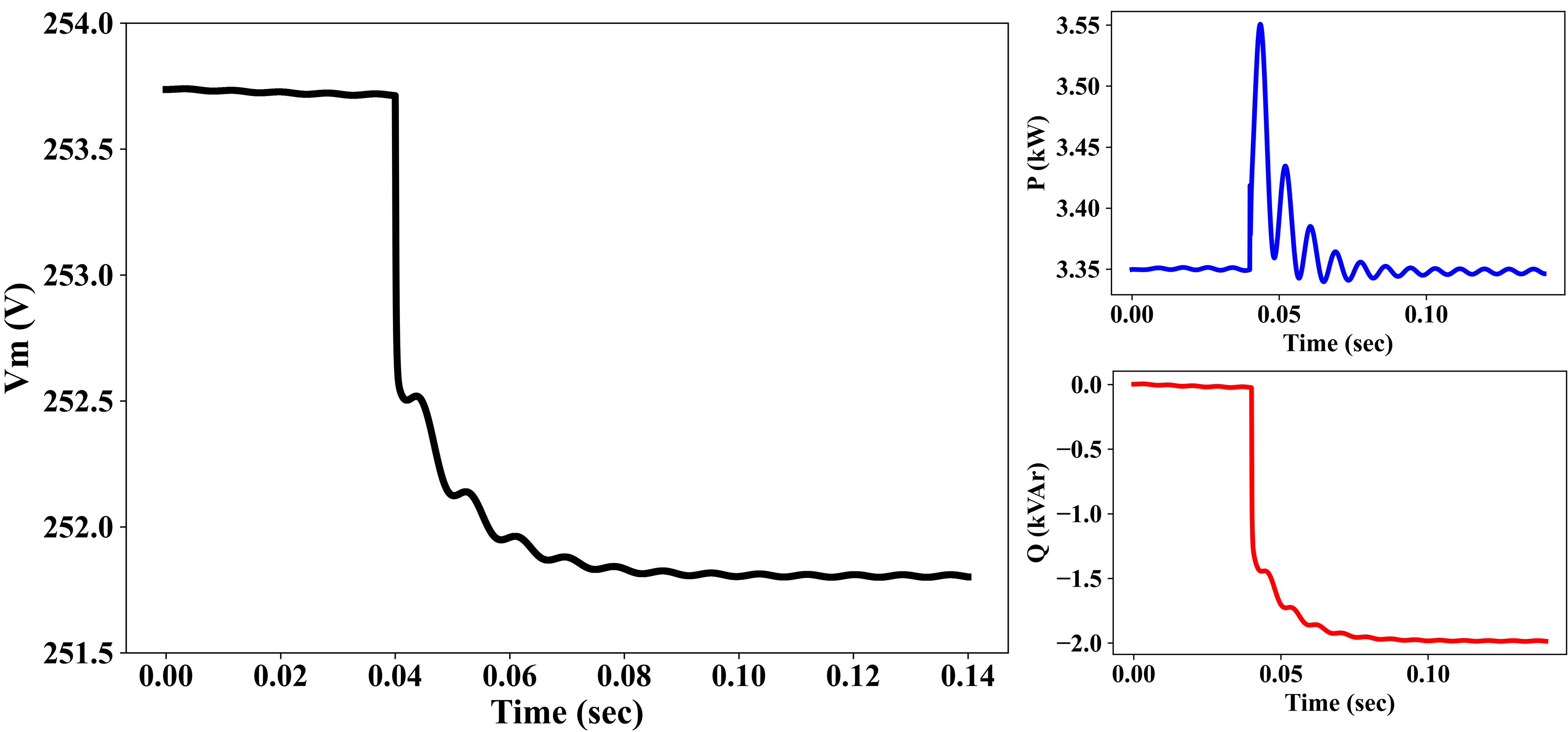}
	\caption{Voltage, active and reactive power at the bus node with the 44th PV smart inverter corresponding to Volt-VAr control enabled at 0.04s in the over-voltage scenario}\vspace{-0.1in}
	\label{fig_over}
\end{figure}

\begin{figure}
	\centering
	\includegraphics[width=0.45\textwidth]{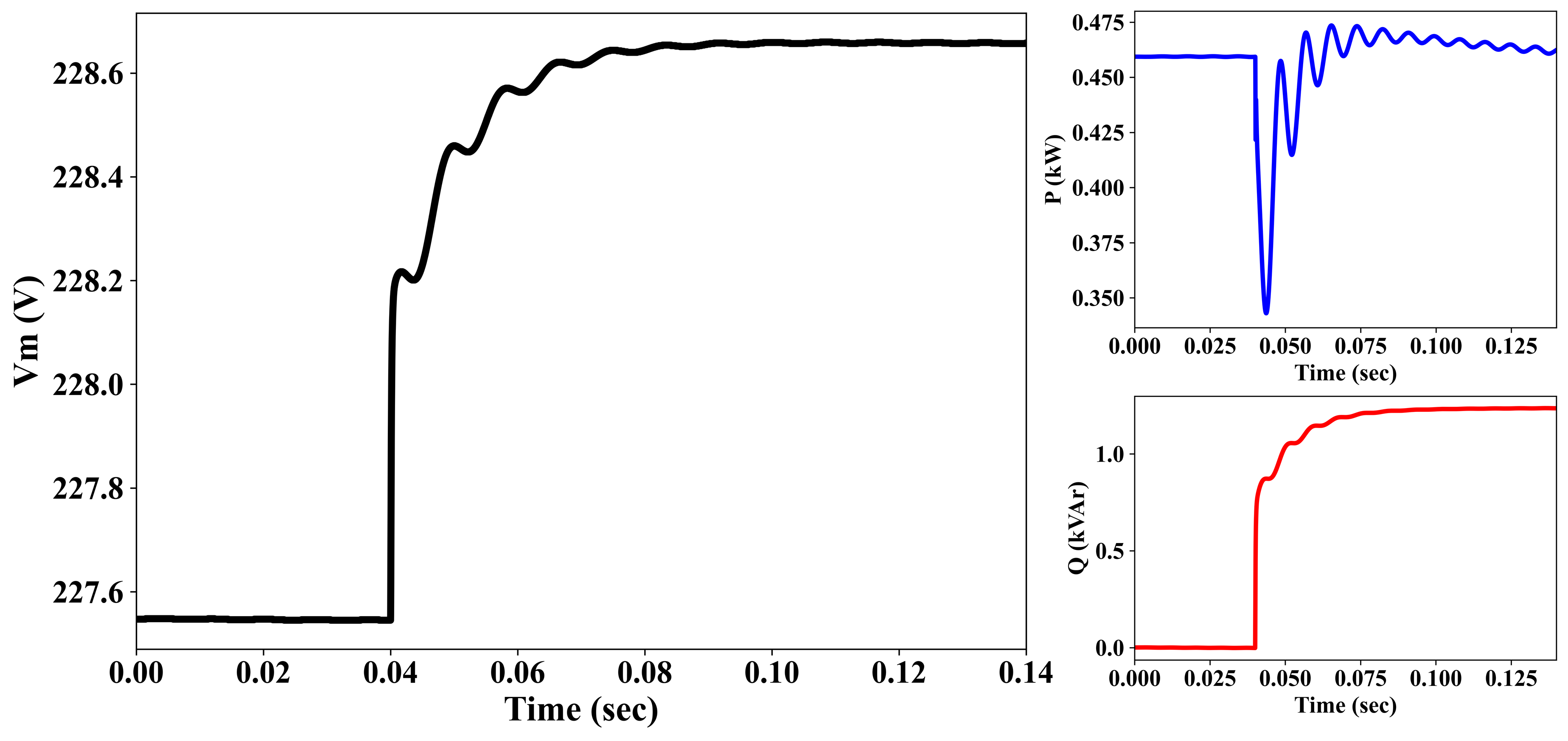}
  \vspace{-0.2in}
	\caption{Voltage, active and reactive power at the bus node with the 1st PV smart inverter corresponding to Volt-VAr control enabled at 0.04s in the under-voltage scenario} 
	\label{fig_under}
\end{figure}

The optimally placed PV smart inverters disable the Volt-VAr control at time $t=0.0$s, and Volt-VAr control is enabled at 0.04s under both scenarios in the dynamic studies. Fig. \ref{fig_over} shows the dynamic results of the 44th PV smart inverter in the over-voltage scenario, which has one of the largest voltage violations before enabling Volt-VAr control. It can be observed that a significant voltage drop is induced by the reactive power absorption of the PV smart inverter at time $t=0.04s$. After the transient period, the active power can be maintained at its original value. The reactive power is kept at a higher value corresponding to the Q-V curve settings to maintain the voltage at a lower value. Similarly, Fig. \ref{fig_under} shows the dynamic results of the 1st PV smart inverter in the under-voltage scenario.  These results show that even in a feeder such as the utility partner’s feeder used in this study with PV penetration exceeding $100\%$, it is possible to manage the feeder voltage profile and keep the system stable using only a relatively small number (99 of 767) of optimally placed PV smart inverters to provide Volt-VAr support. 

\section{Conclusion}
In this paper, a two-stage stochastic mixed-integer linear programming model is proposed to determine optimal numbers and locations of PV smart inverters with Volt-VAr control to mitigate under/over voltage conditions while minimizing the active power curtailment of PV units in active unbalanced distribution networks. In the first stage, the upgrading cost of PV smart inverter with Volt-VAr control is minimized, while the second stage minimizes the expected cost of active power curtailment of PV units and considers the detailed model of Q-V curve characteristics of PV smart inverter according to IEEE 1547 standard. Additionally, a detailed dynamic model of PV smart inverters is developed using DLL in OpenDSS to evaluate the distribution system's steady-state and dynamic stability with the obtained optimal locations of the PV smart inverters under different voltage, load, and PV output scenarios. The results illustrate that the optimal location of PV smart inverters with Volt-VAr control can mitigate the worst over and under-voltage conditions of the utility feeder network within the allowable voltage requirement of the system without any PV active power curtailment. The proposed model utilizes the local control mode of the smart inverters without communication requirement with the control center and other PV smart inverters, which avoids possible adverse impacts of communication delays or failures. Also, the simulation results of dynamic transition from inactive Volt-VAr control mode to active Volt-VAr control mode in the optimally located PV smart inverters at the worst voltage condition illustrate that the system remains stable while reactive power output of PV units is adjusted to maintain the system voltage at the normal operation range.

\bibliographystyle{IEEEtran}

\bibliography{IEEEabrv,Task5_ref}

\end{document}